\documentclass[useAMS,usegraphicx]{mn2e}

\newcommand{\Rsolar}{\mbox{$R_{\odot}\,$}}
\newcommand{\Msolar}{\mbox{$M_{\odot}\,$}}

\newcommand{\kms}{\mbox{${\rm km\,s}^{-1}$}}

\title[The mass ratio distribution of DDs.]
{The mass ratio distribution of short period double degenerate stars.}
\author[P. F. L. Maxted, T. R. Marsh and C. K. J. Moran]
       {P. F. L. Maxted$^{1,2}$, T. R. Marsh$^1$ and C. K. J. Moran$^1$ \\
 $^1$University of Southampton, Department of Physics \& Astronomy,
 Highfield, Southampton, S017 1BJ, UK \\
 $^2$Department of Physics, Keele University, Staffordshire, ST5~5BG,
 UK}

\date{Accepted --- Received ---}

\pagerange{\pageref{firstpage}--\pageref{lastpage}}

\pubyear{2001}

\begin{document}

\maketitle

\label{firstpage}
\begin{abstract} 

Short period double degenerates (DDs) are close white~dwarf\,--\,white~dwarf
binary stars which are the result of the evolution of interacting binary
stars. We present the first definitive measurements of the mass ratio for two
DDs, WD\,0136+768 and WD\,1204+450, and an improved measurement of the mass
ratio for WD\,0957$-$666. We compare the properties of the 6 known DDs with
measured mass ratios to the predictions of various theoretical models. We
confirm the result that standard models for the formation of DDs do not
predict sufficient DDs with mass ratios near 1. We also show that the observed
difference in cooling ages between white dwarfs in DDs is a useful constraint
on the initial mass ratio of the binary. A more careful analysis of the
properties of the white dwarf pair WD\,1704+481.2 leads us to conclude that
the brighter white dwarf is older than its fainter companion. This is the
opposite of the usual case for DDs and is caused by the more massive 
white dwarf being smaller and cooling faster. The mass ratio in the sense
(mass of younger star)/(mass of older star) is then 1.43$\pm$0.06 rather than
the value 0.70$\pm$0.03 given previously. 
\end{abstract}


\section{Introduction}

 Short period double degenerates (DDs) are binary stars in which both stars
are white dwarfs. The orbital periods of DDs are hours or days (Maxted \&
Marsh 1999) so the separation of the stars is only a few solar radii.  White
dwarfs are the remnants of giant stars which have radii of hundreds of solar
radii, much larger than the current size of the binary, so there has clearly
been dramatic shrinkage of the binary orbit during the evolution of the binary
star. There are several models for the formation of DDs which have been used
to predict the properties of this group of binary stars. These usually assume
that the most recent episode of orbital shrinkage is due to a common
envelope (CE) phase  in which a red giant star comes into contact with its
Roche lobe and begins to transfer mass to its companion star. This mass
transfer is highly unstable, so a ``common envelope'' forms around the
companion and the core of the red giant. The drag on the companion orbiting
inside the common envelope  leads to extensive mass loss and dramatic
shrinkage of the orbit (Iben \& Livio 1993). If this CE phase happens while
the star is on the first giant branch, the degenerate helium core of the
red giant will be exposed and will appear as a white dwarf of unusually low
mass, i.e., about 0.4\Msolar, c.f., 0.55\,--\,0.6\Msolar for a typical white
dwarf (Bragaglia et~al. 1995). The mass of a white dwarf can be measured
directly from its spectrum by comparing the surface gravity, $\log g$, and the
effective temperature, T$_{\rm eff}$, to cooling models of white dwarfs. Low
mass white dwarfs identified this way are particularly fruitful source of DDs
(Marsh, Dhillon \& Duck 1995).

 The most notable difference between models for the formation of DDs is the
way they treat the formation of the first white dwarf. For example, Iben,
Tutukov \& Yungelson (1997) have predicted the distributions of masses,
periods and mass ratios for DDs using a numerical model of the population of
close binaries which evolve through either two CE phases or an Algol-like
phase (stable mass transfer on a thermal timescale) followed by a CE phase.
Han (1998) has produced a similar model but has also explored how the various
parameters of the model affect the distribution of periods, mass or mass ratio
which are predicted. Han's model also includes enhanced mass loss in a star as
it approaches its Roche limit, an effect which is not incorporated in the
model of Iben, Tutukov \& Yungelson. It has now been fairly well established 
that these models do not succesfully predict the distribution of mass ratios
for DDs. This problem has been explored on a case-by-case basis by Nelemans
et~al. (2000). The size of a red giant is related directly to its core mass so
they were able to show that for three DDs with measured mass
ratios, the standard prescriptions for the first mass transfer phase predict
orbits which are too small. They used a parametric approach to describe the
first mass transfer phase based on a consideration of the angular momentum
balance during this phase, rather than the more usual energy balance arguments
used by Iben et~al. and Han et~al. This parametric approach was incorporated
into a model for the population of DDs by Nelemans et~al. (2001). They were
able to predict a mass ratio distribution which was more consistent with those
observed provided that they assumed that very low mass white dwarfs ($\la
0.3$\Msolar) cool more rapidly than recent models predict. This assumption is
required to avoid the prediction that most binary white dwarfs should have
very low masses - a problem common to many of these models.

 Measuring  mass ratios for DDs is a powerful way to test models of how binary
stars interact. Strong obervational tests of these models are desirable
because many interesting astrophysical phenomena are the result of interacting
binary stars, e.g., black hole binaries, AM CVn binaries, Type~Ia
supernovae, cataclysmic variables and novae. The properties of these objects
have a direct bearing on areas of astronomy other than the study of binary
stars themselves, e.g., the evolution of the properties Type~Ia supernovae
over the history of the Universe is a matter of direct concern when they are
used as standard candles to measure cosmological parameters (Umeda et~al.
1999). 

 Several preliminary estimates of the mass ratios and others parameters for
WD\,0136+768, WD\,0957$-$666 and  WD\,1204+450 have been published elsewhere
(e.g., Moran, Marsh \& Maxted 1999).  The values given here should be used in
preference to those earlier estimates although the conclusions discussed
above, many of which are based on those preliminary estimates, are not
affected in general by the small changes to the values of the mass ratio given
here.

\section{Observations and reductions}
 Observation of WD\,0957$-$666 were obtained with the RGO spectrograph on the
3.9m Anglo-Australian Telescope (AAT) at Siding Spring, Australia. We used a
1200\,line/mm grating with the 82cm camera and a 0.8\,arcsec slit. The detector
used was a TEK charge-coupled device (CCD) with $1024^2 \times  24\,\mu$m
pixels. The resolution of the spectra is 0.7\AA\ and the dispersion is
0.23\AA-per-pixel.

 Observations of WD\,0136+768 and WD\,1204+450 were obtained with the 500mm
camera of the intermediate dispersion spectrograph (IDS) on the 2.5m Isaac
Newton Telescope (INT) on the Island of La Palma and the ISIS spectrograph on
the 4.2m William Herschel Telescope (WHT), also on La Palma. Both
spectrographs were used with 1200\,line/mm gratings, a 1\,arcsec slit and a
TEK CCD. The resolution of the IDS spectra is 0.9\AA\ and the dispersion is
0.39\AA-per-pixel. The resolution of the ISIS spectra is 0.8\AA\ and the
dispersion is 0.40\AA-per-pixel.

\begin{table}
\caption{\label{ObsTable}Journal of observations.}
\begin{tabular}{@{}lllr@{}}
Name & Telescope & Dates & \multicolumn{1}{l}{No. of } \\
     &           &       & \multicolumn{1}{l}{spectra} \\
WD\,0136+768   & WHT & 1995 Jan 21--24 &   7 \\
               & INT & 1995 Jun 19--24 &  10 \\
               & WHT & 1996 Jan 11--14 &  16 \\
               & INT & 1997 Nov  6--10 &  46 \\
WD\,0957$-$666 & AAT & 1996 Mar   1--3 &  31 \\
               & AAT & 1997 Mar 18--21 & 169 \\
WD\,1204+450   & INT & 1998 Feb  8--11 &  32 \\
               & WHT & 1998 May 13--14 &   3 \\
               & WHT & 1998 Jul  8--13 &   8 \\
               & WHT & 2001 Mar  7--8  &   3 \\
\end{tabular}
\end{table}

 The dates of observation for each star are given in Table~\ref{ObsTable}
The observing procedure is very similar in each case. We obtain spectra of our
target stars around the H$\alpha$ line using exposure times of 5--30\,minutes.
Spectra of an arc lamp are taken before and after each target spectrum with
the telescope tracking the star. None of the CCDs used showed any structure in
unexposed images, so a constant bias level determined from a clipped-mean
value in the over-scan region was subtracted from all the images. Sensitivity
variations were removed using observations of a tungsten calibration lamp. The
sensitivity variations along the spectrograph slit are removed using
observations of the twilight sky in the AAT images because the tungsten
calibration lamp is inside the spectrograph. We have occasionally used the
same technique for the WHT and INT spectra, though it makes little difference
in practice whether we use sky images or lamp images to calibrate these
images.

 Extraction of the spectra from the images was performed automatically using
optimal extraction to maximize the signal-to-noise of the resulting spectra
(Marsh 1989).  The arcs associated with each stellar spectrum were extracted
using the profile determined for the stellar image to avoid possible
systematic errors due to tilted arc lines. The wavelength scale was determined
from a polynomial fit to measured arc line positions and the wavelength of the
target spectra interpolated from the calibration established from the
bracketing arc spectra. Uncertainties on every data point calculated from
photon statistics are rigorously propagated through every stage of the data
reduction. 

\section{Analysis}

 In this section we describe how we have measured the mass ratios for
WD\,0957$-$666,  WD\,0136+768 and WD\,1204+450 using a simultaneous fit to
the H$\alpha$ line in all the available spectra for each star.

\subsection{WD\,0136+768}
\begin{figure}
\caption{\label{Pgram0136} The periodogram of the average radial
velocities measured for WD\,0136+768 measured by cross-correlation.}
\includegraphics[width=0.45\textwidth]{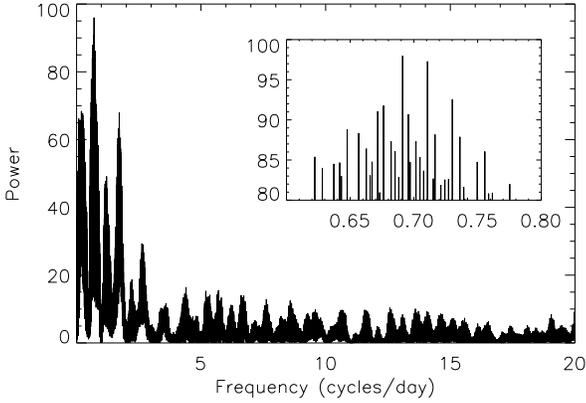}
\end{figure}

 The spectra clearly show two sharp cores to the H$\alpha$ line separated by
up to about 150\,km\,s$^{-1}$ with one core slightly deeper than the other. To
determine an approximate value of the orbital period, we cross correlated the
spectra against a single Gaussian profile with a full-width at half minimum
(FWHM) of 200\,km\,s$^{-1}$ and measured the position of the peak. This gives
an average of the radial velocities of the two stars weighted towards the star
with the deeper core. We then calculated the periodogram of the results over
100\,000 frequencies between 0.01 and 20 cycles per day. The periodogram is
shown in Fig.~\ref{Pgram0136}. All the periodograms in this paper show the
natural logarithm  of the probability ratio P(D$|$C+S,$f$)/P(D$|$C), where
P(D$|$C+S,$f$) is the probability of obtaining the data given that it the
sampling of a signal composed of a sine wave with frequency $f$ plus a
constant, and (P$|$C) is probability of obtaining the data given that it the
sampling of a constant signal, i.e., the exponent of equation (A4) discussed
in Marsh, Dhillon \& Duck (1995). The orbital frequency is clearly near 0.7
cycles per day. We then calculated the periodogram of the results over 20\,000
frequencies between 0.6 and 0.8 cycles per day, also shown in
Fig.~\ref{Pgram0136}. There are two peaks of similar significance in this
range at 0.69112 and  0.71063 cycles per day.

\begin{figure}
\caption{\label{RV0136Fig} The approximate radial
velocities measured for WD\,0136+768 measured by least-squares fitting of
fixed Gaussian profiles. The solid lines shows least-squares fits of sine
functions to either the filled circles or the unfilled circles.}
\includegraphics[width=0.45\textwidth]{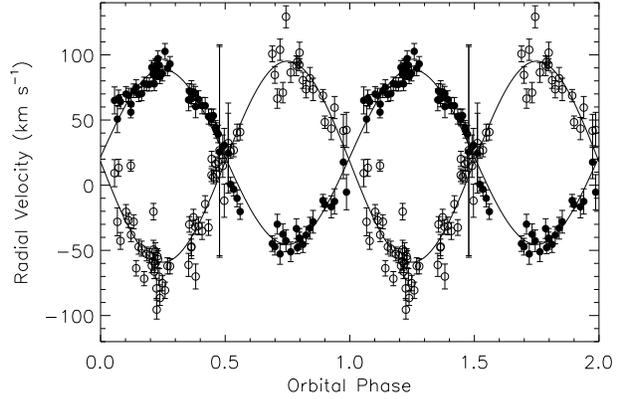}
\end{figure}

 In order to determine approximate values for the projected orbital velocities
of the stars and to identify the correct orbital frequency, we measured
approximate radial velocities for both stars using a least-squares fit of two
Gaussians profiles with fixed widths of 1.2\AA\ and of fixed depths of
10\,percent of the local continuum. A fixed quadratic function was used to
account for the shape of the spectrum near the cores and only data within
500\,km\,s$^{-1}$ of the rest wavelength of H$\alpha$ was included in the
least-squares fit. After assigning the measured the radial velocities to the
correct star for each of the trial frequencies, it was clear from the quality
of a least-squares fit of a sine wave to the data  that 0.71063 cycles per day
is the correct orbital frequency. The fit to these data is shown in
Fig.~\ref{RV0136Fig}.

\begin{table}
\caption{\label{Fit0136Table} Parameters of the final least-squares
fit to the spectra of WD\,0136+768. The full-width at half-minimum (FWHM) and
depths (D) of each of the Gaussian profiles used to model the H$\alpha$ line
are also given. The number of pixels included in the fit is $N$, other symbols
are defined in the text. }
\begin{tabular}{@{}lr@{}}
 {\bf Orbit} \\
 HJD$(T_0)-2450000$    &  760.8533     $\pm$   0.0034    \\
 $P (d)$                 &   1.407221   $\pm$   0.000009  \\
 $\gamma_1(\kms)$ &  22.3 $\pm$ 0.6 \\
 $K_1(\kms)$      &  67.4 $\pm$ 0.8 \\
 $\gamma_2(\kms)$ &  15.0 $\pm$ 1.2 \\
 $K_2(\kms)$      & $-$84.8 $\pm$ 1.8  \\
 {\bf Star 1} \\
  FWHM$_1$ (\AA) & 77.4  $\pm$   1.9  \\
  D$_1$          & 0.150 $\pm$   0.003 \\
  FWHM$_2$ (\AA) & 17.7 $\pm$   1.1 \\
  D$_2$          & 0.052 $\pm$   0.004 \\
  FWHM$_3$ (\AA) &  5.3 $\pm$   0.5 \\
  D$_3$          & 0.039 $\pm$   0.004 \\
  FWHM$_4$ (\AA) & 0.80 $\pm$   0.06\\
  D$_4$          & 0.177 $\pm$   0.008 \\
 {\bf Star 2} \\
  FWHM$_1$ (\AA) & 33.6  $\pm$   1.1 \\
  D$_1$        & 0.098 $\pm$   0.004 \\
  FWHM$_2$ (\AA) &  7.0  $\pm$   0.4 \\
  D$_2$        & 0.047 $\pm$   0.003 \\
  FWHM$_3$ (\AA) &  0.68 $\pm$  0.14 \\
  D$_3$        & 0.082 $\pm$   0.011 \\
 {\bf Continuum} \\
               & 0.958 $\pm$ 0.003 \\
 {\bf Fit} \\
 N             &   43985 \\
 $\chi^2$      & 51897.64 \\
 Reduced $\chi^2$ &     1.181 \\
\end{tabular}
\end{table}

 To measure the radial velocities of the two components more precisely we used
a simultaneous least-squares fit to all the spectra of two model profiles, one
for each star, in which the position of each model profile is predicted from
its time of observation, $T$, using the equation $\gamma_i +
K_i\sin(2\pi(T-T_0)/P)$, where $i=$1 or 2. We define $T_0$ such that star 1
has the deeper H$\alpha$ core and is closest to the observer at time $T_0$.
The projected orbital velocity of star 1 is $K_1$ and its apparent mean
velocity is $\gamma_1$ and similarly for star 2. Note that $\gamma_1 \ne
\gamma_2$ because the apparent mean velocity is the sum of the radial velocity
of the system and the gravitational redshift of each star, and this second
quantity is different for each star.  Each model profile is the sum of a
number of Gaussian profiles with independent widths and depths but with the
same mean. In this way we are able to determine the shape of the two profiles
and the parameters of the two circular orbits, including  $T_0$ and $P$,
directly. The resolution of each spectrum is included by convolution of the
final model profile for each spectrum with a Gaussian function of the
appropriate width and the effects of smearing due to orbital motion are also
included.

\begin{figure*}
\caption{\label{Trail0136} From left-to-right, trailed spectrograms of the 
data, the fit and the residuals for WD\,0136+768. The spectra are shown in
temporal order from bottom-to-top and are displayed as a greyscale. 
The residuals are displayed over a range of $\pm$20\,percent.}
\includegraphics[angle=270,width=0.95\textwidth]{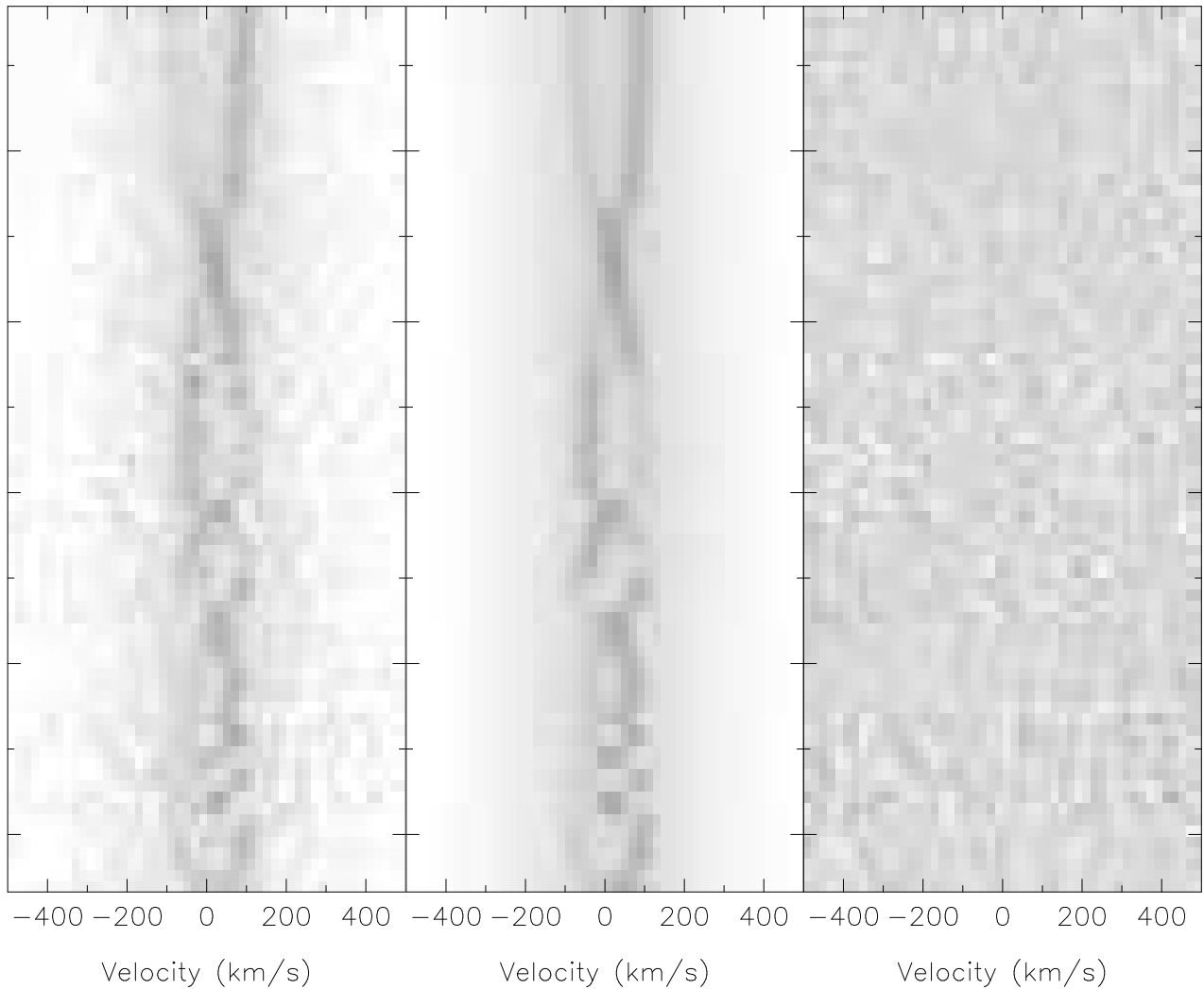}
\end{figure*}

There are many free-parameters in this fitting process so we used a series of
least-squares fits in which first the profile shapes were fixed while the
parameters of the orbit were varied and then {\it vice versa}, until we had
established values for all the parameters which were nearly optimal. Only data
within 5000\,km\,s$^{-1}$ of the rest wavelength of H$\alpha$ is included in
the least-squares fit. The spectra were normalized using a linear fit to the
contiuum either side of the H$\alpha$ line prior to fitting. We used four
Gaussian profiles to model the broad wings of the H$\alpha$ line and the core
of star with the deeper core and three Gaussian profiles for the other star. A
polynomial is also included in the fitting process to allow for smooth,
asymmetric features in the profile. For the final least-squares fit the
parameters of the profile shapes and the orbit were all varied independently.
The reduced $\chi^2$ value for this fit was rather high because of small
changes from spectrum-to-spectrum due to inaccuracies in the normalization
process. We corrected for these small offsets by subtracting the mean of the
residuals over the fitting regions from each spectrum. We then repeated the
fitting process and obtained the results given in Table~\ref{Fit0136Table}. In
fact, this renormalization has a negligible effects on the results. The
trailed spectrograms of the data, the fit and the residuals are shown in
Fig.~\ref{Trail0136}. 

\begin{figure*}
\end{figure*}

\subsection{WD\,0957$-$666}

 Bragaglia et~al. (1990) first noted the variable radial velocity of
WD\,0957$-$666 and the absence of any spectral features due to a main-sequence
companion in the optical or near-infrared spectrum. They concluded that
WD\,0957$-$666 is a double degenerate with an orbital period of 1.15\,d.
Moran, Marsh \& Bragaglia (1997) used the AAT data from 1996
described above to show that the correct orbital period is 1.46\,h. The core
of the H$\alpha$ line from the fainter white dwarf in this binary is visible
in these spectra. This lead to a mass ratio measurement of $q=1.15\pm0.10$. We
obtained additional data of WD\,0957$-$666 with the AAT during 1997 to improve
the measurement of the mass ratio. 

\begin{figure*}
\caption{\label{Trail0957} From left-to-right, trailed spectrograms of the 
data, the fit and the residuals for WD\,0957$-$666 after coadding spectra over
a range of orbital phases. 
The residuals are displayed over a range of $\pm$20\,percent.}
\includegraphics[width=0.75\textwidth,angle=270]{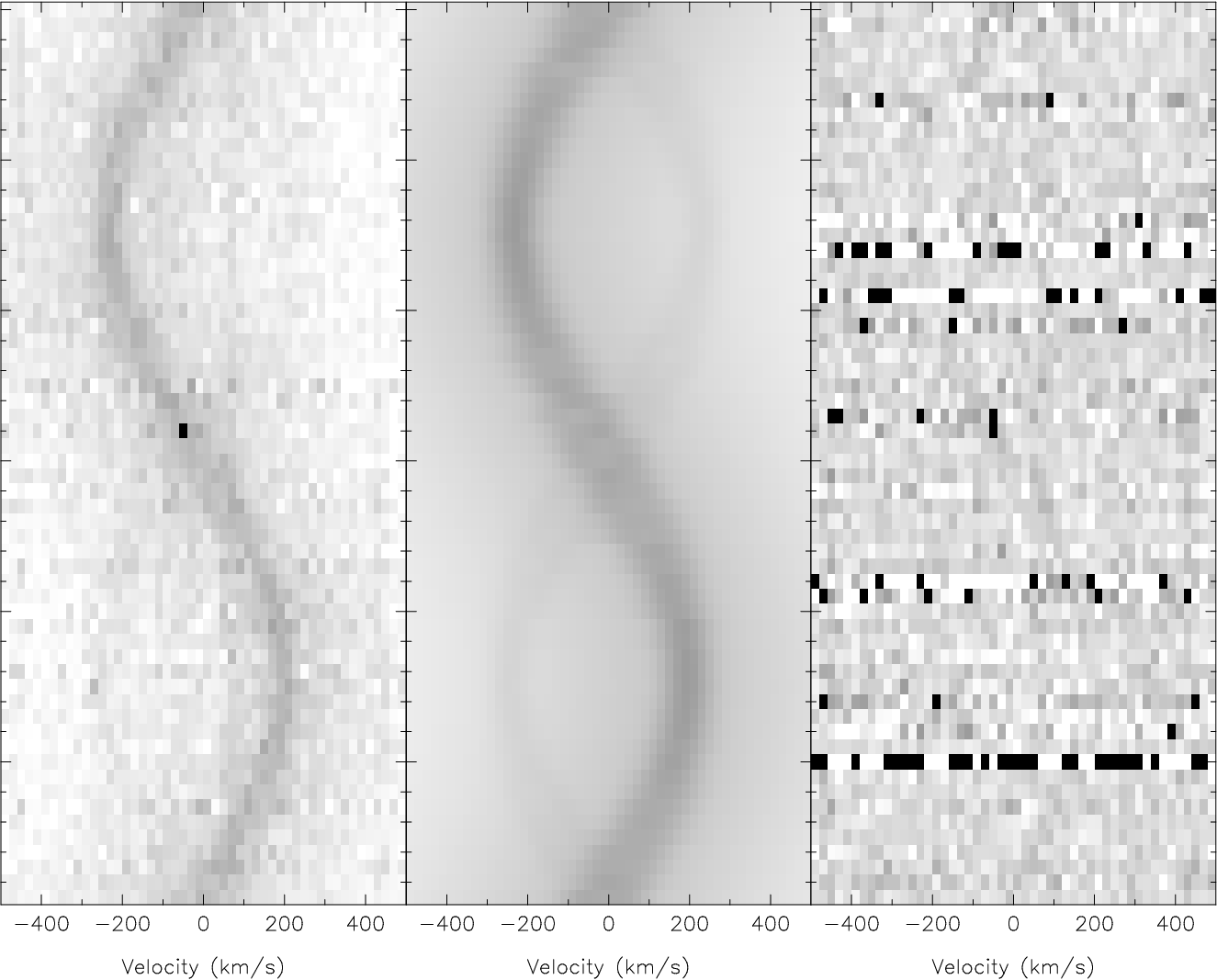}
\end{figure*}

\begin{table}
\caption{\label{Fit0957Table} Parameters of the final least-squares
fit to the spectra of WD\,0957$-$666. The full-width at half-minimum (FWHM) and
depths (D) of each of the Gaussian profiles used to model the H$\alpha$ line
are also given. The number of pixels included in the fit is $N$, other symbols
are defined in the text. The model profiles are convolved with a Gaussian of
FWHM 0.7\AA\ to account for the instrumental resolution.}
\begin{tabular}{@{}lr@{}}
 {\bf Orbit} \\
 HJD($T_0$-2450000)(\kms) &  145.3794 $\pm$   0.0001  \\
 $P$(d) &  0.06099312 $\pm$   0.00000002 \\
 $\gamma_1$(\kms) & $-$18.4 $\pm$   0.8 \\
 $\gamma_2$(\kms) & $-$7.4  $\pm$   3.6 \\
 $K_1$(\kms)      &  218.4 $\pm$   1.1 \\
 $K_2$(\kms)      & $-$246.3 $\pm$   5.0 \\
 {\bf Star 1} \\
 FWHM$_1$(\AA)  & 36.3 $\pm$   5.5 \\
 D$_1$     & 0.100 $\pm$   0.045 \\
 FWHM$_2$(\AA)  &  21.1 $\pm$   8.6 \\
 D$_2$     & 0.052 $\pm$   0.035 \\
 FWHM$_3$(\AA)  &  9.6 $\pm$   2.2 \\
 D$_3$     & 0.039 $\pm$   0.016 \\
 FWHM$_4$(\AA)  &  2.08 $\pm$   0.09 \\
 D$_4$     & 0.162 $\pm$   0.005 \\
 {\bf Star 2} \\
 FWHM$_1$(\AA)  &  27.4 $\pm$   2.1 \\
 D$_1$     & 0.027 $\pm$   0.002 \\
 FWHM$_2$(\AA)  &  1.6 $\pm$   0.2 \\
 D$_2$     & 0.041 $\pm$   0.004 \\
 {\bf Continuum} \\
               & 0.951 $\pm$ 0.007 \\
 {\bf Fit} \\
 N             &   74445 \\
 $\chi^2$      &   78698.4 \\
 Reduced $\chi^2$ &     1.057 \\
\end{tabular}
\end{table}

\begin{table}
\caption{\label{Fit1204Table} Parameters of the final least-squares
fit to the spectra of WD\,1204+450. The full-width at half-minimum (FWHM) and
depths (D) of each of the Gaussian profiles used to model the H$\alpha$ line
are also given. The number of pixels included in the fit is $N$, other symbols
are defined in the text. }
\begin{tabular}{@{}lr@{}}
 {\bf Orbit} \\
 HJD($T_0$-2450000)(\kms) & 1003.789  $\pm$   0.004   \\
 $P$(d) &  1.602663   $\pm$   0.000016   \\
 $\gamma_1$(\kms) & 33.2    $\pm$   1.3 \\
 $\gamma_2$(\kms) & 38.7    $\pm$   1.6 \\
 $K_1$(\kms)      &   99.6 $\pm$   2.2 \\
 $K_2$(\kms)      & $-$86.7  $\pm$   2.6 \\
{\bf Star 1} \\
 FWHM$_1$(\AA)  & 116.7 $\pm$   6.2 \\
 D$_1$     & 0.137 $\pm$   0.007 \\
 FWHM$_2$(\AA)  &  49.6 $\pm$   3.1 \\
 D$_2$     & 0.087 $\pm$   0.010 \\
 FWHM$_3$(\AA)  &  12.4 $\pm$   0.7 \\
 D$_3$     & 0.066 $\pm$   0.004 \\
 FWHM$_4$(\AA)  &  1.71 $\pm$   0.08 \\
 D$_4$     & 0.113 $\pm$   0.004 \\
 {\bf Star 2} \\
 FWHM$_1$(\AA)  &  30.3$\pm$   1.8 \\
 D$_1$     & 0.067 $\pm$   0.005 \\
 FWHM$_2$(\AA)  &  6.5 $\pm$   0.7 \\
 D$_2$     & 0.037 $\pm$   0.004 \\
 FWHM$_3$(\AA)  &  1.24 $\pm$   0.14 \\
 D$_3$     & 0.076 $\pm$   0.005 \\
 {\bf Continuum} \\
           & 1.078 $\pm$ 0.003 \\
 {\bf Fit} \\
 N             &   27965 \\
 $\chi^2$      &   28658.0 \\
 Reduced $\chi^2$ &     1.026 \\
\end{tabular}
\end{table}

\begin{figure}
\caption{\label{Pgram1204} The periodogram of the radial
velocities measured from the deeper H$\alpha$ core of WD\,1204+450. The
full vertical scale in shown in the upper panel and a restricted vertical
range is shown in the lower panel to emphasise the most prominent peaks.}
\includegraphics[width=0.45\textwidth]{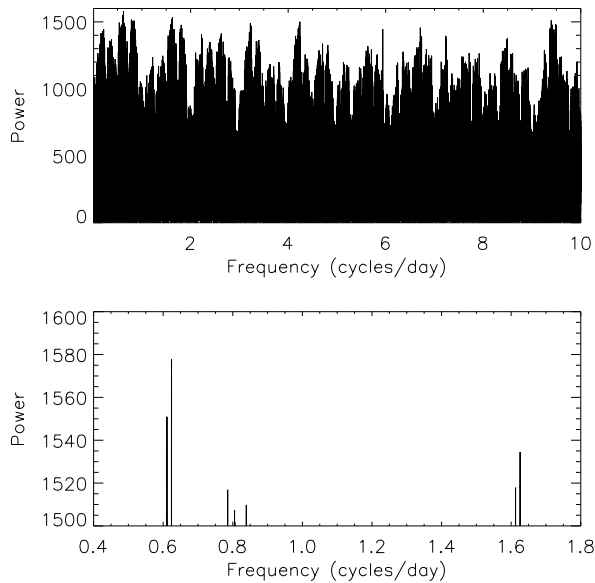}
\end{figure}
\begin{figure}
\caption{\label{RV1204Fig} The radial 
velocities measured from the H$\alpha$ line of WD\,1204+450. The velocities of
the deeper core are plotted with filled circles, the other core with unfilled
symobls. The solid lines show sine wave fit by
least-squares to the data for each core.}
\includegraphics[width=0.45\textwidth]{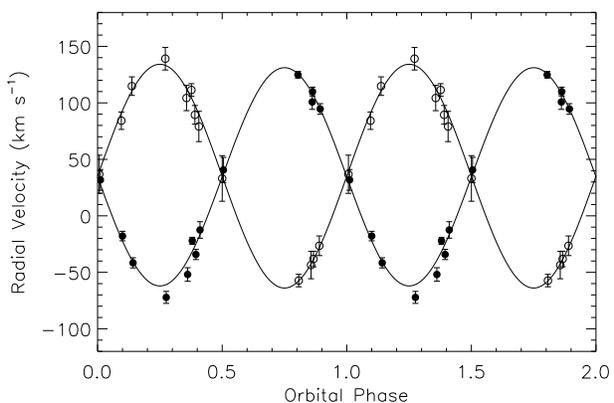}
\end{figure}
\begin{figure*}
\caption{\label{Trail1204} From left-to-right, trailed spectrograms of the 
data, the fit and the residuals for WD\,1204+450. The spectra are shown in
temporal order from bottom-to-top and are displayed as a greyscale. 
The residuals are displayed over a range of $\pm$20\,percent.}
\includegraphics[width=0.75\textwidth,angle=270]{Trail1204.ps}
\end{figure*}

We have used all the spectra of WD\,0957$-$666 from 1996 and 1997
with exposure times of 300s to 600s. We normalized the spectra using a linear
fit to the continuum either side of the H$\alpha$ line. We used the radial
velocites predicted by the ephemeris and circular orbit fit of Moran, Marsh \&
Bragaglia  to apply a shift our 1997 data. Visual inspection of the trailed
spectrogram shows that ephemeris is reliable, so we proceeded directly to a
simultaneous least-squares fit to all the spectra using the same method we
employed for WD\,0136+768.  We found that the quality of the fit is much
improved by using three Gaussians for star 2, rather than two Gaussians as
Moran, Marsh \& Bragaglia used.   Only data within 2000\,km\,s$^{-1}$ of the
rest wavelength of H$\alpha$ is included in the least-squares fit. The optimum
values determined by least-squares for the width and depth of each Gaussian
and the parameters of the circular orbits are  given in
Table~\ref{Fit0957Table}. The trailed spectrograms of the data, the fit and
the residuals are shown in Fig.~\ref{Trail0957}.

\subsection{WD\,1204+450}  
 WD\,1204+450 is a low mass white dwarf (Bergeron, Saffer \& Liebert 1992)
which was first identified as a possible binary by Saffer, Livio \& Yungelson
(1998). We observed WD\,1204+450 with the INT to determine its orbital period.
These spectra show two narrow cores to the H$\alpha$ line, one slightly deeper
than the other, moving in anti-phase by $\sim$100\kms\ over 6--7 hours. This
clearly shows that WD\,1204+450 is a double degenerate with a period of $\sim
1.5$ days but the quality and phase coverage of the INT data is such that it
is difficult to determine the exact orbital period from these data alone,
particularly since it is hard to identify the deeper core in some of the
spectra. We have therefore obtained data with the WHT at slightly better
resolution and higher signal-to-noise which has enabled us to identify an
unambiguous orbital period as follows. 

 We first normalized all the spectra using a low-order polynomial fit to the
continuum either side of the H$\alpha$ line and formed the average of those
spectra taken within 2 hours of each other improve their signal-to-noise. We
used a least-squares fit to a WHT spectrum in which the cores are clearly
distinguished to create a model line profile for each star composed of 4
Gaussian profiles for the star with the deeper core and 3 for the other. We
then  measured the radial velocity of both components in the 18 spectra
available by varying the position of the cores only in a least-squares fit.
The  least-squares fit was done twice for each spectrum, once with the deep
core on the left and once on the right. By comparing the quality of these two
fits, both from the chi-squared value and by-eye, we found 13 spectra in which
the deeper core could be clearly identified or where the velocities of the two
cores are identical to within their uncertainties. The periodogram of the
velocities measured from the deeper core calculated over 100\,000 frequencies
between 0.1 and 10 cycles per day is shown in Fig.~\ref{Pgram1204}. The data
are sparsely sampled so the periodogram is complex, but there is a single
unambiguous peak at an orbital frequency of 0.6240 cycles/day. The periodogram
of the radial velocities for the other core gives the same result. The
measured radial velocities and a sine wave fit by least-squares to the data
for each core are shown in Fig.~\ref{RV1204Fig}. The next most prominent set
of peaks near 1.6 cycles/day are not compatible with the observations taken
over a continuous sequence of 6$\frac{1}{2}$ hours seen in
Fig.~\ref{Trail1204} starting with a conjunction about 1/3 of the way up the
figure and ending near quadrature just over half way up the plot. If the
period were near (1/1.6)d these data would cover almost half an orbit and we
we then expect to see a second conjunction, which is clearly not the case.

 To measure the radial velocities of the two components more precisely we used
a simultaneous least-squares fit to all the spectra, i.e.,  prior to forming
the average spectra,  of two model profiles, one for each star, in which the
position of each model profile is predicted from its time of observation, $T$,
using the equation $\gamma_i + K_i\sin(2\pi(T-T_0)/P)$, as before.  Only data
within 5000\,km\,s$^{-1}$ of the rest wavelength of H$\alpha$ is included in
the least-squares fit. The optimum values determined by least-squares for the
width and depth of each Gaussian  and the parameters of the two circular
orbits are given in Table~\ref{Fit1204Table}. The trailed spectrograms of the
data, the fit and the residuals are shown in Fig.~\ref{Trail1204}.

\section{Discussion}

\subsection{Masses, temperatures and ages}
 In Table~\ref{QTable} we list the mass ratio $q = |K_2|/K_1=M_1/M_2$, where
star 1 has the deeper core, for WD\,0136+768, WD\,0957$-$666 and WD\,1204+450 
based on the values of $K_1$ and $K_2$ measured above. We also list
measurements of $q$ for three other double degenerates where the narrow
H$\alpha$ core of both stars has been measured. Also given in
Table~\ref{QTable} are the orbital period, $P$, the ratio of the depths of the
H$\alpha$ cores, $l_{\rm H\alpha}$, the masses of the two stars, $M_1$ and
$M_2$, estimates of the effective temperatures of the stars, $T_1$ and $T_2$,
and an estimate of the cooling ages of the white dwarfs, $\tau_1$ and
$\tau_2$. We have remeasured $l_{\rm H\alpha}$ for WD\,1704$+$481.2
from one of the WHT spectra described in Maxted et~al. (2000) in which the two
cores are clearly resolved using a least-squares fit of multiple Gaussians in
which the width of the narrow Gaussians assigned to each core are constrained
to have the same width. We also give published mass estimates for the stars
based on their temperatures and surface gravities measured from the optical
spectrum compared to cooling models for white dwarfs. Where only a single mass
from the combined spectrum has been published, we assign individual masses
consistent with the mass ratio such that the luminosity weighted mean mass
equals the published mass. We take the value of $l_{\rm H\alpha}$ as an
estimate of the luminosity ratio since the additional error introduced by this
approximation is small compared to the typical uncertainties on the mass
estimates ($\pm$0.05\Msolar).

  WD0135$-$052 is the only binary in Table~\ref{QTable}  for which separate
effective temperatures have been published for the the two stars. In order to
calculate the effective temperatures of the two stars in the other binaries,
we compared values of $l_{\rm H\alpha}$ and the effective temperature measured
from the combined spectrum to the values predicted by the cooling models of
Benvenuto \& Althaus (1999) as follows. We used linear interpolation between
their grid of models to calculate the radii of the stars as a function of
$T_1$ and $T_2$ for white dwarfs of the appropriate mass. We used the surface
brightness values and V$-$R colours tabulated by Bergeron, Wesemael \&
Beauchamp (1995) combined with the radius and effective temperature calculated
from cooling models to calculate the luminosity ratio in the R-band. We then
multiplied this ratio by the ratio of H$\alpha$ core depths estimated from the
effective temperature to get a grid of $l_{\rm H\alpha}$ values. The core
depth is  taken from a polynomial fit to the depth of the H$\alpha$ core
versus effective temperature measured from Fig.~3 of Koester \& Herrero
(1988). 

\begin{figure}
\caption{\label{T0136Fig} The regions of the $T_1$--$T_2$ plane where the
values of $l_{\rm H\alpha}$ (solid lines) and the mean
effective temperature (dashed lines) predicted from the models of Benvenuto \&
Althaus (1999) are consistent with the observed values for WD\,0136+768 for
models of the appropriate white dwarf mass. }
\includegraphics[width=0.45\textwidth]{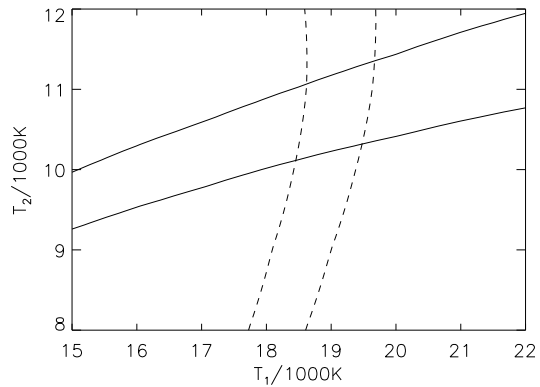}
\end{figure}

 We then assumed that the effective temperature measured from the combined
spectrum is the mean of $T_1$ and $T_2$ weighted by the luminosity on the
V-band. The best estimates of $T_1$ and $T_2$ are then given by intersection
of the region of the $T_1$--$T_2$ plane where the predicted value of $l_{\rm
H\alpha}$ is the consistent with the observed value  and the region where the
mean effective temperature is consistent with its observed value. An example
is given in Fig.~\ref{T0136Fig} for the case of WD\,0136+768. From this figure
we see that the uncertainties in the effective temperatures derived are at
least 500K. Given the approximations and assumptions used to derive these
values, the true uncertainty is probably larger than this, particularly for
fainter companions in binaries with large luminosity ratios. Nevertheless, we
can estimate the cooling age of the stars given their masses and effective
temperatures from the models of  Benvenuto \& Althaus (1999). The
uncertainties on the ages are large  since they depend not only on the
uncertain mass and effective temperature estimates, but also on the unknown
hydrogen layer thickness and other uncertainties in the models. However, we
are most concerned here with the relative ages of the two stars, and this is
less affected by these systematic errors. We estimate the uncertainty in the
relative ages of the stars to be $\sim$50\%.

 The exact age we derive for the stars in WD\,1704$+$481.2 depends on the
cooling model used, but we consistently find that star 1 is older than star
2 whichever model we use. Although star 1 has the deeper H$\alpha$ core, star
2 is is a more massive CO white dwarf so it is smaller and has cooled faster.
This leads to an exception in this case to the general rule that the star with
the deeper H$\alpha$ core has formed  more recently. In all the other cases
the star with the deeper core is the younger of the two, as expected.

\begin{table*}
\caption{\label{QTable} Parameters of double degenerates with measured mass
ratios. The mass ratio is $q$, $P$ is the orbital period, $l_{\rm H\alpha}$ is
the ratio of H$\alpha$ core depths,  the masses are $M_1$ and $M_2$, the
effective temperatures are $T_1$ and $T_2$ and the ages are $\tau_1$
and $\tau_2$. See text for a discussion of the  uncertainties on
these quantities.
}
\begin{tabular}{@{}lrrrrrrrrrr@{}}
Name & 
\multicolumn{1}{l}{$q=\frac{M_1}{M_2}$}& \multicolumn{1}{l}{$P$(days)} & 
\multicolumn{1}{l}{$l_{\rm H\alpha}$}& \multicolumn{1}{l}{$M_1$(\Msolar)} &
\multicolumn{1}{l}{$M_2$(\Msolar)} & \multicolumn{1}{l}{$T_1$(K)} &
\multicolumn{1}{l}{$T_2$(K)} & \multicolumn{1}{l}{$\tau_1$(Myr)}&
\multicolumn{1}{l}{$\tau_2$(Myr)} &References. \\
WD\,0135$-$052 &0.90$\pm$0.04 &1.556 & 1.30 & 0.47 & 0.52 & 7470 & 6920 &
   950 & 1300 & 1, 2 \\
WD\,0136$+$768 &1.26$\pm$0.03 &1.407 & 2.54 & 0.47 & 0.37 & 
   18500 &10500 &150 & 600 & 0, 5 \\
WD\,0957$-$666  &1.13$\pm$0.02 &0.061 & 5.14 & 0.37 & 0.32 &
    30000&11000& 25 & 350 &0, 6 \\
WD\,1101$+$364   &0.87$\pm$0.03 &0.145 & 1.13 & 0.29 & 0.33 & 
  15500 & 12000 & 135 & 350 & 3, 5 \\
WD\,1204$+$450  &0.87$\pm$0.03 &1.603 & 2.05 & 0.46 & 0.52 & 
 31000 & 16000 & 40 & 120 & 0, 5 \\
WD\,1704$+$481.2 &0.70$\pm$0.03 &0.145 & 1.60 & 0.39 & 0.54 & 
 9000& 10000 & 725 & 705 &0, 4 \\
\hline
\multicolumn{11}{l}{0.\,This paper; 1.\,Saffer, Liebert \& Olszewski (1988); 
2.\,Bergeron  et~al. (1989); 3.\,Marsh (1995); 5.\,Bergeron, Saffer \& 
Liebert}\\
\multicolumn{11}{l}{(1992); 6.\,Moran, Marsh \& Bragaglia (1997)}\\
\end{tabular} \\
\end{table*}

\subsection{Comparison to models of DD formation}

 A full comparison of the observed parameters of DDs to all the available
models is beyond the scope of this paper. In this section we re-iterate that
the observed mass ratios are not compatible with ``standard'' models of DD
evolution and show that the cooling ages derived above can be a  useful test
of the feasibilty of models of DD formation.

\begin{figure}
\caption{\label{NelFig} Measured orbital periods, $P$, and mass ratios, $q$,
for DDs compared to the model A2 of Nelemans et~al. (upper panel) and to a
model for ``standard'' common evelope evolution (lower panel). The model
predictions are shown as grey-scale images where darker areas correspond to
regions of the $q-\log P$ plane where the models predict more DDs.}
\includegraphics[width=0.45\textwidth]{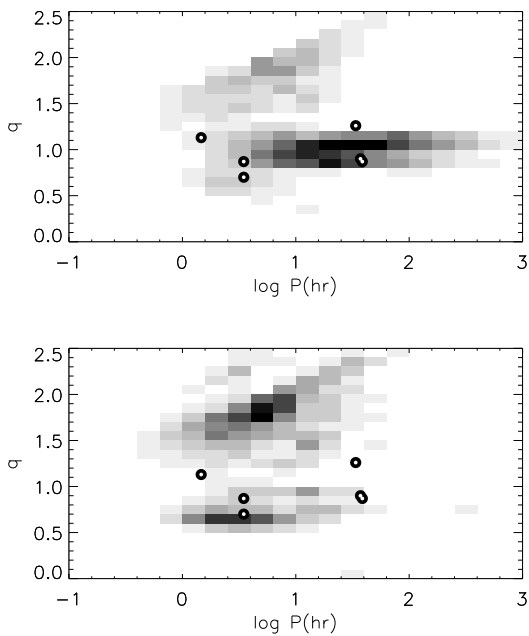}
\end{figure}

 In Fig.~\ref{NelFig} we compare the observed distribution of mass ratios and
orbital periods to the distribution predicted by two different  models taken
from Nelemans et~al. (2001), shown as grey-scale images. These predictions
include selection effects such as a magnitude limit in the observed sample and
the relative luminosity of the two stars. The mass ratio in this case is
$q=m/M$, where $m$ is the mass of the brighter star and $M$ is the mass of its
companion, so the mass ratio of WD\,1704+481.2 is 0.70 according to this
definition. We see that models which assume two phases of common-envelope
evolution (lower panel) do not predict sufficient DDs with mass ratios near
$q\approx 1$. There is better agreement with the distribution of mass ratios
and periods predicted by the model of Nelemans et~al. which uses a
non-standard parametrization of the first mass transfer phase. However, the
number of measured mass ratios is far too small to allow for a more detailed
discussion of the relative merits of different cooling models, prescriptions
for mass loss, initial distribution of periods and masses, etc.

 The non-standard parametrization of the first mass transfer phase used by
Nelemans et~al. (2001) arises from the case-by-case  discussion of the
observed properties of DDs by Nelemans et~al. (2000) described earlier. 
They describe scenarios for the formation of WD\,0136+768, WD\,1101+364 and
WD\,0957$-$666 based on their proposed model of the first mass transfer phase.
These scenarios were devised to explain the observed periods, masses and mass
ratios of these DDs. One further test of the feasibility of scenarios which
was not expoited in the case of WD\,0136+768 and WD\,1101+364 was the relative
cooling ages of the two white dwarfs. The first mass transfer phase occurs
when the more massive star fills its Roche lobe as a result of its rapid
expansion at the end of its main-sequence lifetime. This mass transfer phase
is assummed not to influence the evolution of the less massive star, so the
second mass transfer phase occurs when that star also fills its Roche lobe
after its main-sequence lifetime. The time taken for the mass transfer phases
is negligible compared to these main-sequence lifetimes, so if the two stars
were formed at the same time, we find that their main-sequence lifetimes,
$\tau_{\rm MS,1}$ and $\tau_{\rm MS,2}$, are related to the ages of the white dwarfs,
$\tau_{\rm WD\,1}$ and $\tau_{\rm WD\,2}$ by 
\[\tau_{\rm MS,1} - \tau_{\rm MS,2} \approx \tau_{\rm WD\,2}-\tau_{\rm WD\,1}\]
 where star 1 is the initially less massive star. Since the main-sequence
lifetime of a star is strongly dependent on its mass, the observed value of
$\tau_{\rm WD\,2} - \tau_{\rm WD\,1}$ is a constraint on the initial mass ratio
of the binary, e.g., a larger value of $\tau_{\rm WD\,2} - \tau_{\rm WD\,1}$
implies a larger difference in $\tau_{\rm MS,1} - \tau_{\rm MS,2}$ which in
turn implies a larger difference in the initial masses, i.e., a more extreme
initial mass ratio.
 
 We have used the analytical formula of Hurley et~al. (2000) to calculate the
main-sequence lifetime of a star based on its initial mass for a range of
metallicities. For WD\,0136+768 Nelemans et~al. propose initial masses of
2.25\Msolar\ and 2.12\Msolar\ so we find $\tau_{\rm MS,1} - \tau_{\rm MS,2}
\sim 150$\,Myr. This is rather short compared to the difference in cooling ages
of $\tau_{\rm WD\,2} - \tau_{\rm WD\,1}\sim 450$\,Myr. This suggests that the
initial masses of these stars differed more than proposed by Nelemans et~al.
The main-sequence lifetime is very sensitive to the initial mass and the
measurements of the white dwarf masses and cooling ages are quite uncertain so
it is not possible at this stage to draw more quantitative conclusions. It may
even be the case that the scenario for the formation of WD\,0136+768 proposed
by Nelemans et~al. is consistent with the observed properties of this binary
given the fairly large uncertainties involved. The same cannot be said in the
case of WD\,1101+364. In this case, the initial masses of 1.75\Msolar\ and
2.30\Msolar\ imply a difference in cooling ages of $\sim 800$\,Myr, which is far
too long compared to the difference in cooling ages of $\sim 215$\,Myr. Given
the relatively small difference in cooling ages seen in this DD, it would
appear that a feasible scenario for its formation requires that the initial
masses of the stars are more similar to one another than proposed by  Nelemans
et~al.  The scenario proposed by Nelemans et~al. for the formation of
WD\,0957$-$666 includes a phase of core helium burning for the more massive
star. The observed difference in cooling ages does not constrain the initial
mass ratio directly in this case, but Nelemans et~al. did show that the
relative cooling ages of the two white dwarfs is consistent with the scenario
that they propose in this case.

 The cooling models for helium white dwarfs are uncertain so one should be
careful when drawing conclusions based on quantitative results based on them
as we have done for WD0136+768 and WD1101+364. Following a suggestion by 
a referee, we also considered the models of Driebe et~al. (1998). These
cooling models cover masses of 0.179\Msolar to 0.414\Msolar, so we were able
to apply them to the less massive component of WD0136+768, for example. These
models suggest that this white dwarf is about 1.7\,Gyr old, much older than the
prediction of the Benvenuto \& Althaus models. This greater age is the result
of  a thick hydrogen envelope used in these models.  The ages of the stars in
WD\,1101+364 are increased to 600\,Myr and 1.4\,Gyr when the models of  Driebe
et~al. are used. Although the difference in the ages is exactly as expected
for the scenario outlined by Nelemans et~al., the cooling models for these
very low mass white dwarfs show hydrogen shell flashes. During these shell
flashes the radius of the star increases to $\sim 10\Rsolar$, which would
result in a further common-envelope phase and loss of some fraction of the
thick hydrogen envelope. It is not clear that the models of Driebe et~al. are
applicable to WD\,1101+364 in this case.

\section{Conclusion}

 We have determined the mass ratios and orbital periods of the three double
degenerate stars WD\,0136+768, WD\,1204+450 and WD\,0957$-$666 by measuring
the  radial velocities of both stars from the sharp cores of the H$\alpha$
line. We have compared the measured mass ratios and orbital periods for these
DDs and three others to two models for the formation of DDs, including
selection effects, from Neleman's et~al. (2001). We confirm the result that
standard models for the formation of DDs do not predict suffucient DDs with
mass ratios near 1. We have also shown that the observed difference in cooling
ages between white dwarfs in DDs is a useful constraint on the initial mass
ratio of the binary.

\section*{Acknowledgments}
 PFLM was supported by a PPARC post-doctoral grant. CKJM was supported by a
PPARC postgraduate studentship. The William Herschel Telescope and the Isaac
Newton Telescope are operated on the island of La Palma by the Isaac Newton
Group in the Spanish Observatorio del Roque de los Muchachos of the Instituto
de Astrofisica de Canarias.

\label{lastpage}
\end{document}